\documentclass[runningheads]{llncs}
\usepackage[T1]{fontenc}
\usepackage{graphicx}

\usepackage[labelfont=bf]{caption}
\usepackage[colorlinks=true,linkcolor=blue,citecolor=blue]{hyperref}

\usepackage{amsmath} 
\usepackage{amsfonts} % Mathesymbole
\usepackage{booktabs} % pretty tables
\usepackage{notoccite}

\begin{document}
\title{Graph Residual Noise Learner Network for Brain Connectivity Graph Prediction}
\titlerunning{Graph Residual Noise Learner Network}

\author{Oytun Demirbilek\index{Demirbilek, Oytun}\inst{1}\thanks{Corresponding author: oytun1996@gmail.com. {*} Equal contribution.} \and
Tingying Peng\index{Peng, Tingying}\inst{2,*} \and
Alaa Bessadok\index{Bessadok, Alaa}\inst{2,*}}

\authorrunning{O. Demirbilek et al.}

\institute{Faculty of Computer and Informatics, Istanbul Technical University, Istanbul, Turkey \and Helmholtz AI, Helmholtz Munich, Neuherberg, Germany}

\maketitle              % typeset the header of the contribution

\begin{abstract}
%%%%%%%% 2000 characters version
%%%submitted
A morphological brain graph depicting a connectional fingerprint is of paramount importance for charting brain dysconnectivity patterns. Such data often has missing observations due to various reasons such as time-consuming and incomplete neuroimage processing pipelines. Thus, predicting a target brain graph from a source graph is crucial for better diagnosing neurological disorders with minimal data acquisition resources. Many brain graph generative models were proposed for promising results, yet they are mostly based on generative adversarial networks (GAN), which could suffer from mode collapse and require large training datasets. Recent developments in diffusion models address these problems by offering essential properties such as a stable training objective and easy scalability. However, applying a diffusion process to graph edges fails to maintain the topological symmetry of the brain connectivity matrices. To meet these challenges, we propose the Graph Residual Noise Learner Network (Grenol-Net), the first graph diffusion model for predicting a target graph from a source graph. Its two core contributions lie in (i) introducing a graph diffusion model that learns node-level noise for accurate denoising (ii) introducing a node-based diffusion function to better maintain the topological structure of brain graphs. Our Grenol-Net is composed of graph convolutional blocks, which first learn the source embeddings and second, a set of fully connected layers assisted with a positional encoding block that predicts the nodes with noise level $t-1$ in the target domain. We further design a batch normalization block that learns the target distribution at diffusion timestep $t$ and operates an element-wise subtraction from the predicted nodes with noise level $t-1$. Our Grenol-Net outperformed existing methods on the morphological brain graph extracted from cortical measurements of the left and right hemispheres separately and on three distinct datasets from multiple cohorts.

\end{abstract}
\section{Introduction}
A morphological brain graph represents cortical similarities between pairwise Regions of Interest (ROIs) using structural T1-weighted MRI \cite{mahjoub2018brain}, providing insights into how different brain regions' morphology may influence each another. Studies have shown that morphological connectivities are crucial for exploring brain abnormalities in neurological diseases \cite{xu2021morphological,mahjoub2018brain}. However, acquiring complete connectomic datasets often faces challenges in reality due to the high costs of medical scans and time-consuming neuroimage processing pipelines \cite{fischl2012freesurfer}. To tackle this challenge, graph convolution frameworks based on generative adversarial network (GAN) have been proposed to predict missing brain graphs or to infer structural from functional brain graphs\cite{bessadok2019hierarchical,gurbuzrekik2020,zhang2022predicting,bessadok2022graph}. However, these approaches may suffer from generator hallucination due to the GAN mode collapse problem, leading to a lack of diversity in the predicted graphs despite the large training set. Alternatively, \cite{gurbuzrekik2020} proposed a brain graph generative framework based on edge graph convolutional network (NNConv) \cite{gilmer2017neural} to generate a representative brain graph template of a population of brain multigraphs with a shared neurological state, such as autism. While this method avoids GAN's problems as it relies solely on graph convolutional learning blocks, it is not specifically designed for predicting a morphological target brain graph from a source graph. Consequently, there is a need in the field of network neuroscience for more accurate and dedicated generative models for predicting a missing morphological brain graph from an existing one.

Recently, diffusion models \cite{sohl2015deep,ho2020denoising} have emerged as promising solutions to the data generation problem, demonstrating remarkable progress in generating different types of medical modalities \cite{kazerouni2023diffusion}. The key success of diffusion models largely comes from the noise diffusion process that introduces diversity in the generated data, making it less prone to mode collapse issues \cite{dhariwal2021diffusion}. Despite recent progress in these methods for medical image generation, existing graph diffusion models are mostly designed for predicting proteins or molecular structures, which have different topologies from brain graphs \cite{liu2023generative}. Thus, these graph diffusion frameworks cannot be used for \emph{brain graph generation} tasks. So far, we have identified only one brain graph synthesis work using a diffusion model \cite{rajadhyaksha2023diffusion}. This work combines a graph U-net with a diffusion model in a single framework to predict high-resolution brain graphs from low-resolution ones, where resolution refers to the number of ROIs considered in the brain graph. Although promising, this model faces one significant challenge: the diffusion process proposed in the work does not preserve the node-wise topological properties as it operates on the graph edges directly. However, the edges represent the similarity in morphology between ROIs (nodes), which are crucial for understanding the function of each brain region in both healthy and neurological disorder cases \cite{bassett2017network}. Therefore, neglecting the diffusion process on nodes fails to maintain the symmetry of the connectivity patterns in the brain. 

%\footnote{Python package to process and prepare brain graph datasets at \url{https://pypi.org/project/avicortex/}
We propose Grenol-Net\footnote{Grenol-Net code in Python on GitHub at \url{https://github.com/oytundemirbilek/grenol-net/} }, the first graph diffusion framework tailored to predict a target brain graph from source one, effectively learns the topological brain structure given the node features, thus maintaining the connectivity diversity in isomorphic brain graphs (i.e., derived from the same parcellation brain template). Our framework comprises two intricately designed learning blocks, each playing a pivotal role in achieving our objectives. The first block comprises a series of graph convolutional blocks that unravel the complex interplay of connections within the source brain graph. By leveraging a cross-node message passing function, Grenol-Net learns a unique embedding of each ROI, laying a solid foundation for precise prediction. This process is further enhanced by a set of fully connected layers, complemented by a positional encoding block, which orchestrates the prediction of nodes with noise level $t-1$ in the target domain. This integration of graph convolution and positional encoding enables Grenol-Net to capture the intricate nuances of brain graph connectivity with unparalleled accuracy. The second block of Grenol-Net introduces a batch normalization mechanism that not only learns the distribution of the target brain graph at diffusion timestep $t$ but also establishes a residual connection to isolate and recover the noise applied to the target. We offer open access to pre-trained Grenol-Net and its source code, fostering collaborative research efforts within the network neuroscience community to expand upon our work.

\section{Methodology}

\begin{figure}
\centering
\includegraphics[width=\textwidth]{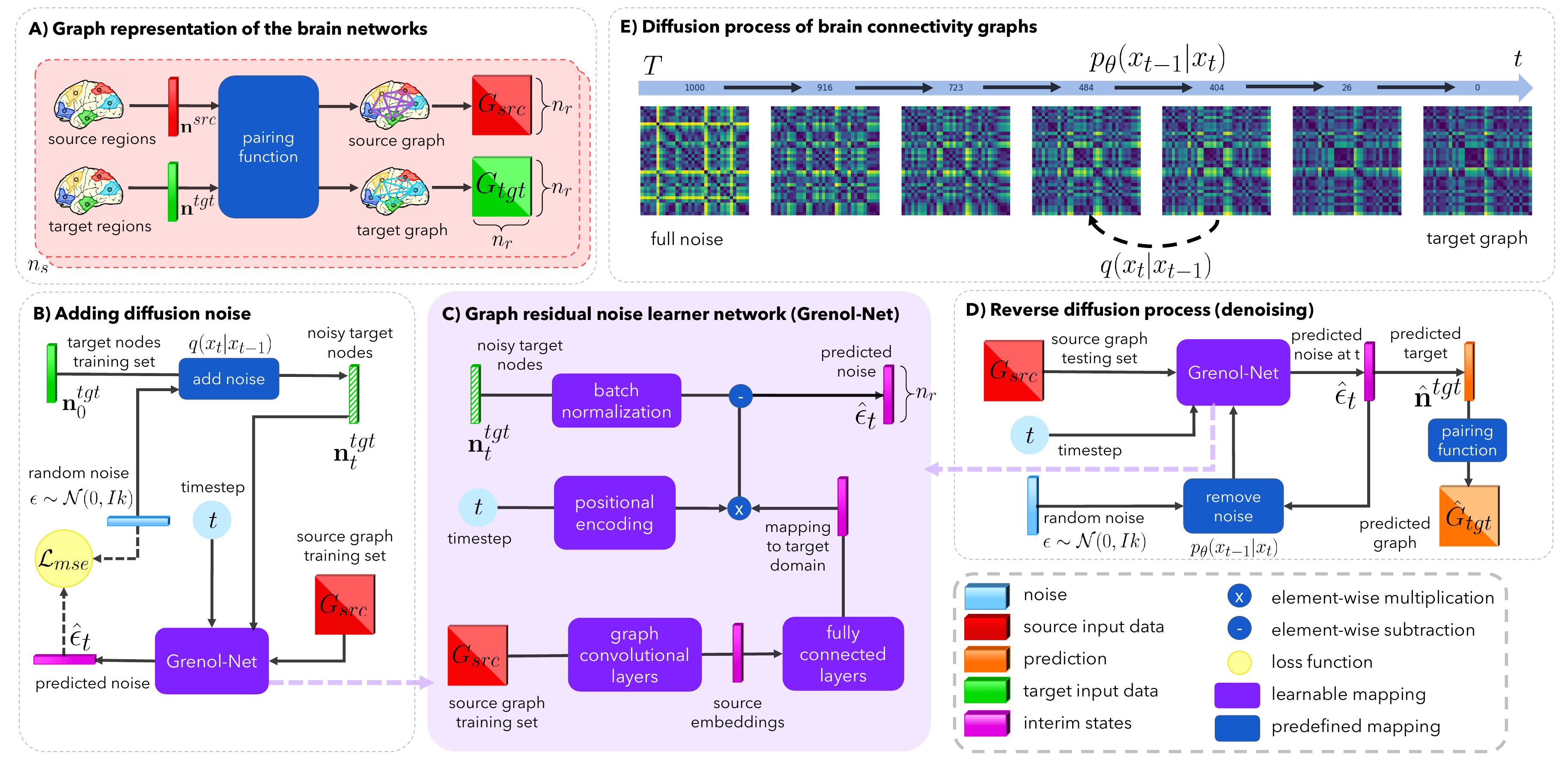}
\caption{\textit{Overview of our Grenol-Net architecture for predicting a target brain graph from source graph by learning conditional denoising process.}}
\label{fig1}
\end{figure}

%%%%%%%%%%%%%%%%%%%%%%%%%%%%%%%%%%%%%%%%%%%%%%%%%%%%%%%%%%%%%

\textbf{A) Brain graph representation.} Let $G(V, E)$ be a fully connected undirected graph where $V$ is the set of vertices (i.e., nodes) and $E$ is the set of edges. In the context of brain graph prediction, our objective is to learn a mapping from source $G_{src}(V_{src}, E_{src})$ to target $G_{tgt}(V_{tgt}, E_{tgt})$. Particularly, we construct edges $e_{i,j}$ each represents a morphological (i.e., specific structural features that represent shape) dissimilarity between two cortical regions of interest (ROI) (i.e., respective node features $n_i$ and $n_j$ of node $i$ and node $j$), and calculated using our \textit{pairing function} (\textbf{Fig.~\ref{fig1}-A}) as following:\begin{equation}\label{eq_pairing}e_{i,j} = \frac{| n_i - n_j |}{n_i + n_j}\end{equation} which yields a symmetric adjacency matrix. Different from \cite{mahjoub2018brain} where edges are calculated as the absolute difference, here we propose our formula to obtain normalized features strictly between 0 and 1 for a better representation of morphological similarities in brain graphs.

%%%%%%%%%%%%%%%%%%%%%%%%%%%%%%%%%%%%%%%%%%%%%%%%%%%%%%%%%%%%%

\textbf{B) Forward diffusion process to add noise to brain graphs.}
Diffusion models are generative models defined as latent variable models \cite{sohl2015deep}. Let $q(x_t)$ be the forward diffusion process for the target data distribution where $t$ is the diffusion timestep. Sampling from this distribution to get the nodes of a target graph is defined as a Markov chain $x_t \sim  q(x_t)$ to gradually add Gaussian noise according to a variance schedule $\beta_1 ... \beta_T$ \cite{ho2020denoising}:
\begin{equation}
q(x_{1:T}|x_0) = \prod _{t=1}^{T} q(x_t|x_{t-1}), \quad q(x_t|x_{t-1}) = \mathcal {N}(x_t; \sqrt{1 - \beta _{t-1}}x_{t-1}, \beta _t Ik)
\end{equation}
where $I$ is an identity matrix, $T$ is the maximum number of diffusion timesteps, and here we introduce $k$ as the standard deviation coefficient for the normal distribution ($\mathcal{N}$). By proposing the $k$ parameter to scale the standard deviation of the noise, we aim to avoid significant data losses between two diffusion timesteps. Authors in \cite{chen2023importance} also highlighted this problem such that the high variance of independent noise destroys an important amount of information in small-sized images due to less redundancy in neighboring values. 
%Moreover, The authors in \cite{chen2023importance} investigated the effects of noise scheduling in detail, and concluded that for the same noise level denoising becomes simpler in bigger images consisting of redundancy in data due to correlation among nearby pixels.
Next, we denote $\alpha _t = 1 - \beta _t$ and $\overline{\alpha }_t = \prod _{s=1}^{t} \alpha _s$ and we diffuse the node features of target graph $G_{tgt}$ as:
\begin{equation} 
\mathbf{n}_t^{tgt} = \sqrt{\overline{\alpha }_t} \mathbf{n}_0^{tgt} + (1 - \overline{\alpha }_t) \epsilon
\end{equation}
where $\epsilon$ is the generated noise, and $\mathbf{n}_t^{tgt}$ is the nodes of the target graph (\textbf{Fig.~\ref{fig1}-B}). We visualize the output noisy graphs in \textbf{Fig.\ref{fig1}-E} by reconstructing their edges with our pairing function in \textbf{Eq.\ref{eq_pairing}}. Authors in \cite{rajadhyaksha2023diffusion} applied a diffusion process that directly operates on the graph edges that fails to maintain the symmetry of the adjacency matrices. Instead, we unprecedentedly propose our novel diffusion methodology for brain connectivity graphs to address this problem.

%%%%%%%%%%%%%%%%%%%%%%%%%%%%%%%%%%%%%%%%%%%%%%%%%%%%%%%%%%%%%

\textbf{C) Graph residual noise learner network.} We aim to predict the added diffusion noise on a target graph with a given diffusion timestep $t$ and guided with the source graph $G_{src}$. To employ the diffusion guidance, authors in \cite{ho2022classifier} investigate the necessity to train a separate classifier. However, we define a simpler network Grenol-Net as $\hat{\epsilon} = \epsilon_{\theta}(\mathbf{n}_t^{tgt}, t, G_{src})$ where $\hat{\epsilon}$ is the predicted noise and $\theta$ is the learnable parameters (weights and biases). First, graph convolutional layers introduced in \cite{gilmer2017neural} (NNConv) of our model learn an embedding vector to represent the source graph $G_{src}$, each graph convolutional layer can be formulated as follows:
\begin{equation}
n^{\prime}_i = \mathbf{\Theta} n_i + \sum_{j \in N(i)} n_j \cdot f_{\mathbf{\Theta}}(e_{i,j})
\end{equation}
where, $n^{\prime}_i$ is the embedding value for node $i$, $N$ is the neighboring function, and $f_{\mathbf{\Theta}}$ is a linear layer with learnable parameters $\mathbf{\Theta}$ (weights and biases). Then, we introduce a group of fully connected layers that map the source embeddings to target embeddings. We condition these target embeddings by adding a sinusoidal position embedding layer inspired by the Transformer \cite{vaswani2017attention} as it was also proposed to represent the diffusion timestep in \cite{ho2020denoising} since the target embeddings ideally represent the noisy target and are dependent on the diffusion timestep $t$. Next, we introduce a batch normalization layer to the noisy node features from a batch of target graphs $\mathbf{n}_t^{tgt}$ to learn the distribution of the target domain. After we obtain the target embedding vector, we subtract it from the batch-normalized noisy node features vector as the target embeddings should be close to the noisy node features of the previous diffusion timepoint. We refer to this set of operations as residual, since it works as a by-pass connection as shown in \textbf{Fig.~\ref{fig1}-C}. Finally, we define our objective function as the standard mean square error $\mathcal{L}_{mse}$ between the generated noise $\epsilon$ and predicted noise $\hat{\epsilon}$ as follows in closed form:
\begin{equation}
\label{eq:5}
\mathcal{L}_{mse} = \mathbb {E}_{t \sim [1,T], \mathbf{n}_0^{tgt} \sim q(\mathbf{n}_0^{tgt}), \epsilon \sim \mathcal {N}(0,Ik)} \Vert \epsilon - \epsilon _{\theta }(\mathbf{n}_t^{tgt}, t, G_{src}) \Vert ^2
\end{equation}

%%%%%%%%%%%%%%%%%%%%%%%%%%%%%%%%%%%%%%%%%%%%%%%%%%%%%%%%%%%%%
\textbf{D) Guided denoising from source to target domain.}
 The reverse process of the diffusion (i.e., denoising) is defined as a Markov chain $p(x_T) = \mathcal{N}(x_T; 0, Ik)$ that gradually removes Gaussian noise, and formulated as \cite{ho2020denoising}:
\begin{equation}
p_{\theta }(x_{0:T}) = p(x_T) \prod_{t=1}^{T} p_{\theta }(x_{t-1}|x_t), p_{\theta }(x_{t-1}|x_t) = \mathcal {N}(x_{t-1}; \mu_{\theta }(x_t, t), \Sigma_{\theta }(x_t, t))
\end{equation}
where $\Sigma_{\theta }$ is a reverse process covariance function approximator and $\mu_{\theta }$ is a reverse process mean function approximator, the result of the analysis in \cite{ho2020denoising}. They also set $\Sigma_{\theta }(x_t, t) = \sigma_t^2I$ and $\sigma_t=\sqrt{\beta_t \frac{1-\overline{\alpha}_{t-1}}{1-\overline{\alpha}_t}}$ as untrained time dependent constants. Specifically, to approximate the forward process posterior variance and mean, we follow a similar procedure by defining $\mu_{\theta}$ and $\sigma_t$. We picked the same $\sigma_t$ in \cite{ho2020denoising}, but we define our reverse process mean function approximator $\mu_{\theta}$ guided by the source graph ($G_{src}$) which is learned by the graph convolutional layers of our Grenol-Net ($\epsilon_{\theta}$):
\begin{equation}
\mu_{\theta}(\mathbf{n}_t^{tgt},t,G_{src}) = \frac{1}{\sqrt{\alpha_t}} \left ( \mathbf{n}_t^{tgt} - \frac{1-\alpha_t}{\sqrt{1-\overline{\alpha}_t}} \epsilon_{\theta}(\mathbf{n}_t^{tgt},t,G_{src}) \right )
\end{equation}
Finally, we can define our complete sampling procedure $x_{t-1} \sim p_{\theta}(x_{t-1}|x_t)$ to approximate the noisy target nodes at the previous diffusion timestep as follows:
\begin{equation}
\mathbf{n}_{t-1}^{tgt} = \mu_{\theta}(\mathbf{n}_t^{tgt},t,G_{src}) + \sigma_t \epsilon
\end{equation}

With our sampling methodology, we first generate a random Gaussian noise and denoise with source guidance and then gradually remove the noise as illustrated in \textbf{Fig.\ref{fig1}-D}. By introducing a guidance procedure for the sampling methodology of the diffusion model, we aim to generate target graphs from source brain graphs using a diffusion model.

\begin{table}[]
\centering
\caption{Detailed information about datasets.} \label{tab1}
\begin{tabular}{@{}ccccccc@{}}
\toprule
\textbf{\begin{tabular}[c]{@{}c@{}}Dataset  \quad \\ name\end{tabular}} &
  \textbf{\begin{tabular}[c]{@{}c@{}}Number of \quad \\ subjects\end{tabular}} &
  \textbf{\begin{tabular}[c]{@{}c@{}}Parcellation \quad \\ software\end{tabular}} &
  \textbf{MRI Scanner } &
  \textbf{\begin{tabular}[c]{@{}c@{}}Field \quad \\ strength\end{tabular}} &
  \textbf{Age group} &
   \\ \midrule
Candishare & 94 & Freesurfer   & GE Signa & 1.5T & 4-17  &  \\ \midrule 
HCP   & 1113 & HCP pipeline & Siemens Skyra & 3T-7T & 22-35 &  \\ \midrule
Openneuro & 42 & Freesurfer   & Philips Intera & 3T & 18-25 &  \\ \bottomrule

\end{tabular}
\end{table}

\section{Experiments and Results}

\subsection{Datasets}
We evaluated our architecture on 3 different datasets: Human Connectome Project (HCP) Young Adult \cite{HCP}, Child and Adolescent NeuroDevelopment Initiative (Candishare) Schizophrenia Bulletin 2008 \cite{frazier2008candishare,kennedy2012candishare}, and Openneuro Cannabis Use Disorders Identification Test (Openneuro) \cite{koenders2016cudit}. The parcellation statistics were already calculated with the HCP pipeline and verified by further quality checks for the HCP dataset. For Openneuro and Candishare, we executed the Freesurfer \cite{fischl2012freesurfer} pipeline (v7.2.0) on Ubuntu 22.04. We picked datasets to validate our model and benchmarks on various dataset groups. Specifically, all our datasets are collected for different studies, age groups, parcellation software, and MRI machines stated in \textbf{Table ~\ref{tab1}}. A node $n_i$ yields one of the \textit{node features}, each derived from regional statistics from cortical parcellations (e.g., mean curvature, cortical thickness, surface area). To parcellate the brain into cortical regions, we used Desikan-Killiany Cortical Atlas \cite{desikan2006automated}, where each subject is parcellated into 34 ROI and left and right hemispheres from their T1-w MRI scan. We constructed $G_{\text{src}}$ based on mean curvature, while $G_{\text{tgt}}$ were derived from cortical thickness. These morphological metrics were chosen for their significance in network neuroscience studies, as evidenced by prior research
\cite{gurbuzrekik2020,bessadok2019hierarchical}. \footnote{Python package to process and prepare brain graph datasets at \url{https://pypi.org/project/avicortex/}}

\subsection{Hyperparameter tuning}
We evaluated Grenol-Net against benchmark methods using 5-fold cross-validation. Each cortical hemisphere data was independently trained and tested. Our model architecture consists of a graph convolutional block with 3 layers of NNConv with convolution size 48 followed by 3 fully connected layers with size: 128. Training employed Adam optimization with weight decay regularization (AdamW), a learning rate of 0.001, weight decay of 0.001, and 500 epochs. We applied cosine beta scheduling for the noise schedule with hyperparameters set to $k = 0.01$ and $T = 100$.

\subsection{Benchmarks}
In our experiment, we compared our model against three benchmarks tailored for isomorphic brain graph prediction: \textit{First}, hierarchical adversarial domain alignment (HADA) \cite{bessadok2019hierarchical}, a GAN-based method for brain graph prediction using leave-one-out (LOO) splitting. \textit{Second}, multi-GCN-based generative adversarial network (MGCN-GAN) \cite{zhang2022predicting}, employing complex loss functions including MSE, Pearson correlation coefficient (PCC), and GAN loss. \textit{Third}, a graph convolutional architecture \cite{gurbuzrekik2020} predicts a single graph from multi-graphs. We adapt the same architecture to predict a target graph from a source graph.

\subsection{Evaluation}

\begin{figure}
\includegraphics[width=\textwidth]{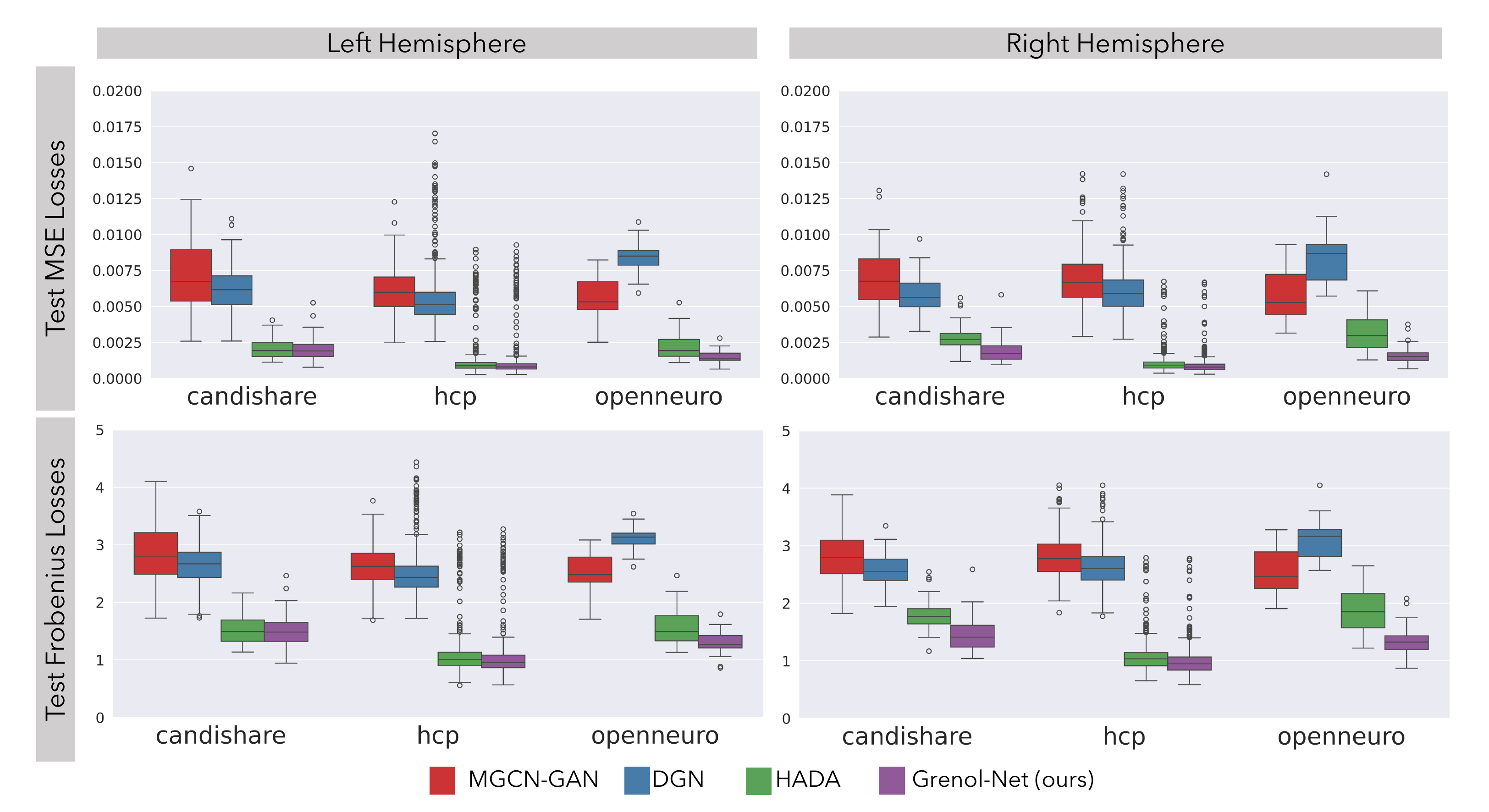}
\caption{\textit{Evaluation in terms of Frobenius distance and mean squared error (MSE) of the respective model MGCN-GAN \cite{zhang2022predicting}, DGN \cite{gurbuzrekik2020}, HADA \cite{bessadok2019hierarchical}, and Grenol-Net separately on different hemispheres.} }\label{fig2}
\end{figure}

To evaluate the accuracy of the predicted target graph, we computed the MSE loss and Frobenius distance loss between the predicted and target adjacency matrix for each subject, separately for both cortical hemispheres. 

\begin{figure}
\includegraphics[width=\textwidth]{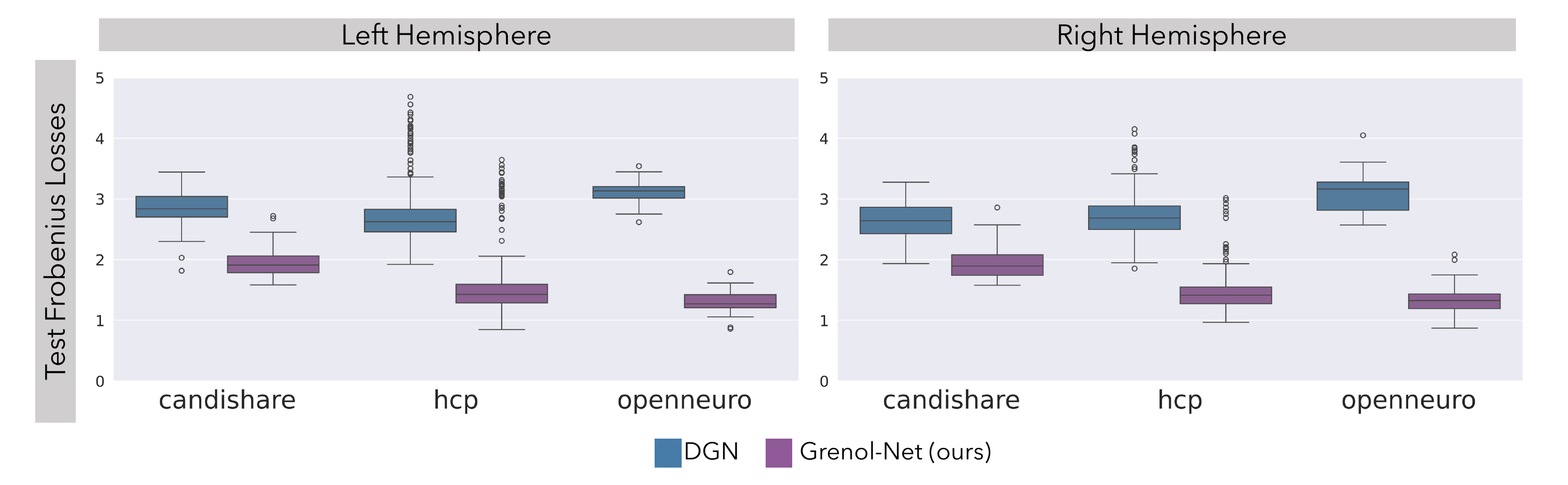}
\caption{\textit{Comparison in terms of Frobenius distance of DGN \cite{gurbuzrekik2020} and Grenol-Net trained on Openneuro dataset and tested on cross-cohort datasets.}}
\label{fig3}
\end{figure}
%{$p<xx$, $∗∗p < 0.01$, $∗∗∗p < 0.001$, Friedmann test with adjusted significance level}). 
Then, we illustrated the results in \textbf{Fig.~\ref{fig2}} for each subject and compared our model with the benchmarks where each model is trained and tested on three datasets separately. Our Grenol-Net significantly outperforms the benchmarks in terms of prediction accuracy across all datasets with $p-value < 0.001$ except for HADA on Candishare's left hemisphere dataset. We showcase our model's superiority over HADA which is a rigid model utilizing the LOO strategy for graph prediction. Likewise, despite the ambiguous inference mode of MGCN-GAN, our model consistently outperforms it across the left and right hemispheres of three datasets. Moreover, MGCN-GAN is trained by optimizing a combination of loss functions: MSE loss, PCC loss, and GAN loss. We illustrate in \textbf{Fig.~\ref{fig2}} that our Grenol-Net trained with simple loss function \textbf{Eq.~\ref{eq:5}} outperforms MGCN-GAN trained with its complex loss function. We further showcase the superiority of our model over DGN, a fully GNN-based generative model. Hence we affirm that our graph diffusion-based model stands out as the superior choice for predicting brain graphs. We further reported results in \textbf{Fig.~\ref{fig3}} of Grenol-Net being trained and validated with solely 21 subjects and tested on 21 others from Openneuro (internal test set) and 47 test subjects from Candishare and 557 test subjects from hcp (cross-cohort test sets). Remarkably, our model still performs well on other cohorts. This demonstrates the breadth of our residual noise learner in accurately training on a very limited dataset. Thus, we can affirm the robustness of Grenol-Net in training and generalizing to diverse cohorts, showcasing its capability to effectively learn from a limited dataset while maintaining high performance across different subject groups. We ran our experiments using a single NVIDIA GeForce RTX 3060 with 6GB memory and measured the inference time of our model on one subject as 0.1816 seconds on average. To conclude, Grenol-Net's remarkable performance not only highlights its ability to surpass the limitations of benchmark models through its diffusion-based property but also underscores its efficiency in inference, as evidenced by its ability to process individual predictions in mere seconds. This versatility positions Grenol-Net as an ideal solution for a wide range of isomorphic graph prediction tasks.

\section{Conclusion}

Our graph diffusion model effectively addresses the morphological brain network prediction challenge. Future research could focus on enhancing the model's explainability and interpretability to deepen insights into brain connectivity dynamics. Additionally, while applicable to one-target prediction tasks, the computational cost remains a challenge for multi-target predictions. Nevertheless, our simple yet efficient framework holds the potential for transforming brain graph prediction, with future endeavors aiming to develop more efficient diffusion models for multi-target prediction while emphasizing interpretability to advance neuroscience understanding.

\section{Acknowledgement} This project has been funded by the Humboldt Postdoctoral Research Fellowship supporting A. Bessadok. However, all scientific contributions made in this project are owned and approved solely by the authors.

\bibliographystyle{splncs04}
\bibliography{Paper-0009}

\end{document}